\documentclass[prd,aps, preprint,  nofootinbib,superscriptaddress,11pt]{revtex4-1}
\usepackage{amsmath}
\usepackage{amssymb}
\usepackage{epsfig}
\usepackage{subfigure}
\usepackage{slashed}
\usepackage{axodraw}

\begin{document}

\title{Electroweak Baryogenesis And The Fermi Gamma-Ray Line}

\author{Jonathan Kozaczuk}
\email{jkozaczu@ucsc.edu}\affiliation{Department of Physics, University of California, 1156 High St., Santa Cruz, CA 95064, USA}

\author{Stefano Profumo}
\email{profumo@scipp.ucsc.edu}\affiliation{Department of Physics, University of California, 1156 High St., Santa Cruz, CA 95064, USA}\affiliation{Santa Cruz Institute for Particle Physics, Santa Cruz, CA 95064, USA} 

\author{Carroll L. Wainwright}
\email{cwainwri@ucsc.edu} \affiliation{Department of Physics, University of California, 1156 High St., Santa Cruz, CA 95064, USA}

\date{\today}

\begin{abstract}
\noindent Many particle physics models attempt to explain the 130 GeV gamma-ray feature that the Fermi-LAT observes in the Galactic Center.  Neutralino dark matter in non-minimal supersymmetric models, such as the NMSSM, is an especially well-motivated theoretical setup which can explain the line.   We explore the possibility that regions of the NMSSM consistent with the 130 GeV line can also produce the observed baryon asymmetry of the universe via electroweak baryogenesis.  We find that such regions can in fact accommodate a strongly first-order electroweak phase transition (due to the singlet contribution to the effective potential), while also avoiding a light stop and producing a Standard Model-like Higgs in the observed mass range.  Simultaneously, CP-violation from a complex phase in the wino-higgsino sector can account for the observed baryon asymmetry through resonant sources at the electroweak phase transition, while satisfying current constraints from dark matter, collider, and electric dipole moment (EDM) experiments.  This result is possible by virtue of a relatively light pseudoscalar Higgs sector with a small degree of mixing, which yields efficient $s$-channel resonant neutralino annihilation consistent with indirect detection constraints, and of the moderate values of $\mu$ required to obtain a bino-like LSP consistent with the line. The wino mass is essentially a free parameter which one can tune to satisfy electroweak baryogenesis. Thus, the NMSSM framework can potentially explain the origins of both baryonic and dark matter components in the Universe. The tightness of the constraints we impose on this scenario makes it extraordinarily predictive, and conclusively testable in the near future by modest improvements in EDM and dark matter search experiments.

\end{abstract}

\maketitle

\section{Introduction}\label{sec:intro}

In the search for signatures from the annihilation (or the decay) of dark matter particles, a gamma-ray line in the multi-GeV energy range has long been considered a Holy Grail. Given that, in the weakly interacting massive particle (WIMP) paradigm, Galactic dark matter is virtually at rest, the pair annihilation of two particles into a final state consisting of two photons would produce a monochromatic line with an energy exactly corresponding to the particle dark matter mass (or to half its mass in the case of decay). The advent of the Fermi gamma-ray Large Area Telescope (LAT) heralded promise of potentially delivering this smoking gun signal, which would then serve as a beacon for further searches to close in on a well-defined particle dark matter mass.

Despite a null result presented by the LAT collaboration in Ref.~\cite{Abdo:2010nc}, independent scholars analyzed the Fermi data employing optimized signal-to-noise regions, unveiling a tantalizing excess localized around 130 GeV\footnote{Recent re-analyses with reprocessed data using ``Pass 7 Clean'' events put the line at 135 GeV \cite{fermisymposium}, but nothing qualitative changes in the present discussion, where we will assume the line is at 130 GeV.} and originating from regions including the Galactic center \cite{wenigerline,wenigeretal}. Subsequent independent analyses confirmed the original claim, typically attributing an even larger level of confidence to the discovery of a monochromatic line in the Fermi-LAT data from the center of the Galaxy \cite{Su:2012ft}.

Understandably, the discovery of the line spurred a great deal of interest in the community: a feature in the Earth limb photon events at the same energy was found, albeit with a much lower statistical significance \cite{Su:2012ft}; despite significant efforts in pinpointing possible instrumental or environmental effects that could explain the excess (see e.g.\ Ref.~\cite{otherline}), at present the line feature appears statistically significant enough to deserve serious consideration.

From a model-building and phenomenological standpoint, the 130 GeV line poses interesting challenges: with default choices for the dark matter density profile in the Galaxy, the required pair-annihilation cross section for dark matter (at rest, i.e. at ``zero temperature'') into two photons is about $\left<\sigma v\right>_{\gamma \gamma}\sim 10^{-27} cm^3/s$, much larger than would be expected by suppressing by a factor $\alpha^2$ the pair annihilation cross section expected for WIMP thermal production in the early universe. Even more problematic is the absence of a continuum gamma-ray signal accompanying the line in the region where the line is detected. This poses the question of how to suppress final states that would generously produce e.g.\ neutral pions from hadronization showers of strongly interacting particles, or inverse Compton or bremsstrahlung photons from charged leptons.

Simple paradigms for WIMP dark matter fail at explaining the needed features of the 130 GeV line. For example, neutralinos within the minimal supersymmetric extension of the Standard Model (MSSM) feature large suppressions in the pair annihilation into two photons with respect to any other final state, and the required large rate for neutralino pair-annihilation into two photons cannot be accommodated with the right thermal relic abundance \cite{Bergstrom:1997fh}.

%%%%%%%%%%%%%%%%%%%%%%%%%%%%%%%%%%%%%%%%%%%%%%%%%%%
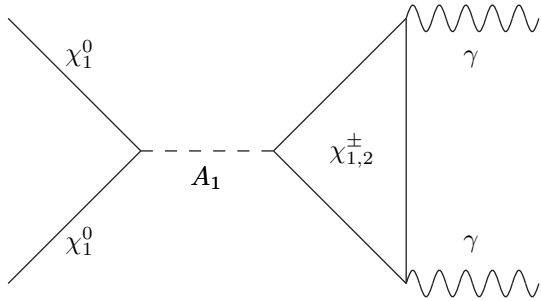
\begin{figure*}[!t]
\begin{center}\begin{picture}(200,100)(0,0)
\Line(0,0)(50,50)
\Line(50,50)(0,100)
\DashLine(50,50)(100,50){5}
\Line(100,50)(150,0)
\Line(100,50)(150,100)
\Line(150,0)(150,100)
\Photon(150,100)(200,100){5}{5}
\Photon(150,0)(200,0){5}{5}
\Text(27,16)[]{$\chi_1^0$}
\Text(27,87)[]{$\chi_1^0$}
\Text(75,40)[]{$A_1$}
\Text(75,40)[]{$A_1$}
\Text(130,50)[]{$\chi_{1,2}^\pm$}
\Text(175,15)[]{$\gamma$}
\Text(175,85)[]{$\gamma$}
\end{picture} 
\end{center}
\caption{\label{fig:diagram}\it\small  The dominant diagram leading to the two-photon pair-annihilation of neutralinos in the NMSSM scenario under consideration in this study.}
\end{figure*}
%%%%%%%%%%%%%%%%%%%%%%%%%%%%%%%%%%%%%%%%%%%%%%%%%%%

A simple extension to the field content of the MSSM, however, allows for an interesting caveat to both shortcomings mentioned above, as first realized in Ref.~\cite{Das:2012ys}: within the next-to-MSSM (or NMSSM, hereafter), an $s$-channel resonant contribution exists to the annihilation cross section arising from the diagram shown in Fig.~\ref{fig:diagram}, where two approximately 130 GeV bino-like neutralinos annihilate into a singlet-like pseudoscalar $A_1$, which then decays into photons via a chargino loop.  For $m_{A_1}\sim 260$ GeV, the process is resonant and the resulting cross-section can easily satisfy $\left<\sigma v\right>_{\gamma \gamma}\sim 10^{-27} cm^3/s$ as required to produce the observed line \cite{wenigerline}.

The NMSSM possesses the interesting additional possibility of naturally realizing a mechanism known as {\em electroweak baryogenesis} to produce the observed baryon asymmetry of the universe (BAU) at the electroweak phase transition (EWPT) (for a recent review, see Ref.~\cite{Morrissey:2012db}). The NMSSM framework, in fact, accommodates tree-level cubic couplings entering the relevant scalar effective potential driving the EWPT needed to produce a sufficiently strongly first-order phase transition (this is in turn needed to prevent wash-out of the generated baryon asymmetry in regions of broken electroweak phase), as realized a long time ago~\cite{Pietroni:1992in, Funakubo:2005pu} and reinforced in recent analyses~\cite{Cheung:2012pg} (see Refs.~\cite{Menon:2004wv,Huber:2006wf} for similar arguments in related models). Additionally, the NMSSM, like the MSSM, possesses enough room to host the level of CP violation needed for baryogenesis while being consistent with constraints from the non-observation of electric dipole moments (EDMs).

In the present study, we argue that the NMSSM can simultaneously accommodate: 
\begin{enumerate}
\item a thermal dark matter candidate that can produce the 130 GeV line while being consistent with constraints from other gamma-ray observations and direct detection searches;
\item a Higgs sector consistent with the recent LHC findings \cite{CMS, ATLAS};
\item a strongly first-order phase transition as needed by electroweak baryogenesis (for which we calculate in detail the effective finite temperature potential);
\item the generation of the observed baryon asymmetry of the universe at the EWPT, while being consistent with constraints from EDMs.
\end{enumerate}

Requiring all four conditions above forces us to very special corners of the theory's parameter space: the goal of our study is not to explore exhaustively the NMSSM parameter space but, rather, to outline the general implications for the theory parameter space of the four requirements above, and to draw predictions from the regions of parameter space that do satisfy these requirements. As a result, we do not concern ourselves with issues of fine-tuning but, rather, we produce a detailed set of predictions that put this framework for the origin of baryonic and dark matter on very testable grounds. At the same time, we provide benchmarks for corners of the NMSSM theory parameter space where all conditions listed above may be fulfilled.

This paper is organized as follows: in Sec.~\ref{sec:nmssm} we outline the NMSSM parameter space, detail the neutralino and Higgs sectors, and discuss the phenomenological constraints we implement; Sec.~\ref{sec:EWPT} discusses the nature of the electroweak phase transition and the constraints that a strongly first-order transition places upon the parameter space; in Sec.~\ref{sec:EWB} we discuss the computation of the baryon asymmetry; we conclude in Sec.~\ref{sec:disc}

\section{A 130 GeV Line in the NMSSM}\label{sec:nmssm}

To begin, we review the NMSSM setup, and show how it is possible to hone in on parameters consistent with the 130 GeV gamma-ray signal and with a broad set of additional phenomenological constraints.  We follow closely the strategy outlined in Refs.~\cite{Das:2012ys, Chalons:2012xf} and consider the simplest incarnation of the NMSSM with a scale-invariant, $\mathbb{Z}_3$-symmetric superpotential: 
\begin{equation} \label{eq:super} W=W_{\rm MSSM}|_{\mu=0}+\lambda \widehat{S} \widehat{H}_u \widehat{H}_d + \frac{\kappa}{3} \widehat{S}^3, \end{equation} 
where hatted quantities denote the corresponding superfields, and where $S$ is a gauge singlet.  The soft supersymmetry-breaking Lagrangian is given by \begin{equation} \label{eq:soft} -\mathcal{L}^{soft}=-\mathcal{L}^{soft}_{\rm MSSM} + m_S^2 \left|S\right|^2 +\left( \lambda A_{\lambda} S H_u H_d + \frac{1}{3} \kappa A_{\kappa} S^3 \right) + {\rm h.c.} \end{equation}  After electroweak symmetry breaking (EWSB), the Higgs and singlet fields obtain vacuum expectation values (vevs) of $\left<H_u\right>\equiv v_u$, $\left<H_d\right>\equiv v_d$, and $\left<S\right>\equiv v_s$.  As in the MSSM, we denote the ratio of the $SU(2)$ Higgs vevs as $\tan\beta\equiv v_u/v_d$.  The singlet vev generates an effective $\mu$-term in the superpotential given by $\mu\equiv \lambda v_s$.  We assume that $\lambda, v_s\in \mathbb{R}$ so that $\mu$ is real and there is no CP-violation at tree level in the Higgs sector.  While CP-violating effects can enter at one-loop from gaugino interactions if we allow $M_{1,2}$ to carry a complex phase, we neglect these contributions when considering radiative corrections to the Higgs sector, since these effects are typically sub-dominant.  The six parameters $\lambda$, $\kappa$, $A_{\lambda}$, $A_{\kappa}$, $\mu$ and $\tan \beta$ then determine the tree-level Higgs spectrum after minimizing the scalar potential and solving for the SUSY-breaking Higgs masses.

At this level, deviations from the spectrum of the MSSM originate from the singlet superfield in the superpotential, and are crucial in order to obtain a neutralino consistent with the 130 GeV gamma-ray signal (without an associated continuum gamma-ray background), with a 125 GeV Higgs, and with successful electroweak baryogenesis.  Specifically, the present set-up contains one each of additional neutral CP-even and CP-odd states which enter into the respective Higgs mixing matrices.  Complete expressions for the various relevant mass matrices in the NMSSM which match our conventions can be found in, e.g., Ref.~\cite{Ellwanger:2004xm}.  

The pseudoscalar mass matrix will be of particular importance; its elements are given, to one-loop order, by \cite{Ellwanger:2004xm} \begin{equation} \label{eq:mA} \begin{aligned} \mathcal{M}^2_{P,11} &=\lambda v_s \left(A_{\lambda}+\kappa v_s\right) \left(\frac{\tan \beta(Q)}{Z_{H_d}}+\frac{\cot \beta(Q)}{Z_{H_u}}\right) \\ \mathcal{M}^2_{P,22}&=4\lambda \kappa v_u(Q)v_d(Q)+\lambda A_{\lambda}\frac{v_u(Q)v_d(Q)}{v_s}-3\kappa A_{\kappa} v_s\\ \mathcal{M}^2_{P,12}&=\lambda \left(\frac{v_u(Q)^2}{Z_{H_d}}+\frac{v_d(Q)^2}{Z_{H_u}} \right)^{1/2}\left(A_{\lambda}-2\kappa v_s\right), \end{aligned} \end{equation}
where $Q$ is the relevant SUSY energy scale; $v_{u,d}(Q)$ and $\tan\beta(Q)$ are the Higgs vevs and $\tan\beta$ at the scale $Q$; and $Z_{H_{u,d}}(Q)$ are wave-function renormalization factors.  The matrix $\mathcal{M}_P$ can be diagonalized to obtain the pseudoscalar mass eigenstates $A_1$ and $A_2$.  As we discuss below, in the present setup $A_1$ must be singlet-like; the state $A_2$ will therefore correspond to an MSSM-like pseudoscalar Higgs boson.

In addition to the new degrees of freedom in the Higgs sector, there is an additional Weyl fermion (the ``singlino", $\widetilde{S}$), corresponding to the fermionic component of the singlet superfield $\widehat{S}$.  This fermionic degree of freedom enters into the neutralino mixing matrix, whose components are given at tree level by \cite{Ellwanger:2004xm} \begin{equation} \label{eq:neutralino} \mathcal{M}_{\chi^0}=\left(\begin{array}{ccccc}M_1 \hspace{.3cm}& 0\hspace{.3cm}&\frac{g_1v_u}{\sqrt{2}}&-\frac{g_1v_d}{\sqrt{2}}\hspace{.3cm}&0\\ .&M_2&\frac{g_2v_u}{\sqrt{2}}&\frac{g_2v_d}{\sqrt{2}}&0\\.&.&0&-\mu&-\lambda v_d\\.&.&.&0&-\lambda v_u \\.&.&.&.&2\kappa v_s \end{array} \right). \end{equation}  Here, we shall consider the case in which the baryon asymmetry is sourced by CP-violation in the higgsino-gaugino sector \cite{Morrissey:2012db}. The masses in Eq.~(\ref{eq:neutralino}) are therefore generically complex-valued.  We will further restrict ourselves to the case of a single complex physical phase, in the wino mass $M_2$, with all other parameters real\footnote{Note that the physical phase we consider here effectively corresponds we to the phase $\phi\equiv{\rm arg}(\mu M_2b^*)$, see e.g.\ Ref.~\cite{Li:2008ez}}.  This results in CP-conservation at tree-level in the Higgs sector.  Since in our construction the LSP is bino-like throughout all of the parameter space we consider, a $CP$-violating phase in $M_1$ would produce large effects on the calculation of the various dark matter properties; we therefore impose $M_1\in \mathbb{R}$.  Eq.~(\ref{eq:neutralino}) is diagonalized by the unitary complex matrix $\mathcal{N}$: \begin{equation} \mathcal{M}_{\chi^0}'=\mathcal{N}^*\mathcal{M}_{\chi^0}\mathcal{N}^{\dagger} \end{equation} and the neutralino masses are given by \begin{equation} \operatorname{diag}\left(m_{\chi_1^0}^2, \hspace{1mm} m_{\chi_2^0}^2, \hspace{1mm}  m_{\chi_3^0}^2, \hspace{1mm} m_{\chi_4^0}^2, \hspace{1mm} m_{\chi_5^0}^2\right)= \mathcal{M}_{\chi^0}'^{\dagger}\mathcal{M}_{\chi^0}'.\end{equation} The five neutralinos are admixtures of $\widetilde{B}$, $\widetilde{W}$, $\widetilde{H}_{u,d}$, and $\widetilde{S}$, the lightest of which will be the lightest supersymmetric particle (LSP) in our setup.  The chargino mass matrix is simply that of the MSSM, again with a possible complex phase in the wino mass entry, yielding the mass eigenstates $\chi_{1,2}^{\pm}$.

Motivated by the lack of a SUSY particle discovery at the LHC, we will assume that all sfermions are heavy\footnote{Note that the authors of Ref.~\cite{Das:2012ys} considered rather light sleptons to account for the possible discrepancy of the muon $g-2$ with the value predicted by the SM.  However, in the present case, such light sleptons can result in large one-loop contributions to the electric dipole moments inconsistent with the constraints discussed in Sec.~\ref{sec:EDMs}, barring cancellations.}, with $m_{sf}\gtrsim 1.5$ TeV.  This effectively decouples them from any processe of interest here.  As a result, to determine the properties of neutralino dark matter, the electroweak phase transition, and the CP-violating sources for electroweak baryogenesis in the present set-up, one must specify the following nine NMSSM parameters:  \begin{equation}  \lambda, \hspace{.3cm} \kappa, \hspace{.3cm}  A_{\lambda}, \hspace{.3cm}  A_{\kappa}, \hspace{.3cm}  \mu, \hspace{.3cm} \tan \beta, \hspace{.3cm} M_1, \hspace{.3cm} \left| M_2\right| \hspace{.3cm} \phi\equiv \operatorname{arg}(M_2). \end{equation} As we argue below, many of these parameters are tightly constrained by the phenomenological and observational constraints we impose, in particular by requiring a 130 GeV gamma ray line from resonant neutralino annihilation consistent with other particle and dark matter searches. 

\begin{figure*}[!t]
\mbox{\hspace*{0cm}\includegraphics[width=0.6\textwidth,clip]{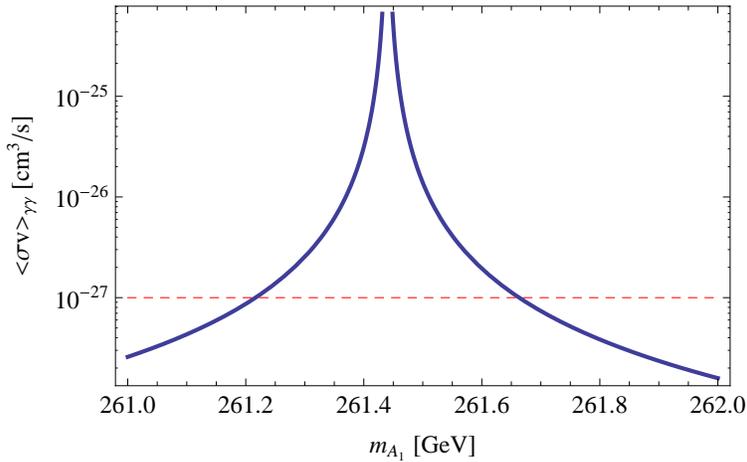}}\caption{\label{fig:cross_section}\it\small  The zero-temperature thermally-averaged cross-section times velocity for neutralino annihilation into two photons as a function of the singlet-like pseudoscalar mass $m_{A_1}$ for the EWPT benchmark point discussed in Sec.~\ref{sec:EWPT}: $\lambda=0.75$, $\kappa=0.45$, $\tan\beta=1.7$, $A_{\lambda}=545$ GeV, $A_{\kappa}=-88$ GeV, $\mu=275.8$ GeV, $M_1=143.5$ GeV, and $M_2=635.5$ GeV.  The red dashed line indicates the lower bound on $\left<\sigma v\right>_{\gamma \gamma}$ required to produce the 130 GeV Fermi line.  Note that decreasing $M_1$ (thereby increasing $\mu$) will narrow down the resonance.}
\end{figure*}

Throughout this study, we will assume that the large required pair-annihilation cross-section into two photons, $\left<\sigma v\right>_{\gamma \gamma}\geq 10^{-27}$ cm$^3/$s, arises from the on-resonance $s$-channel annihilation of neutralinos into $A_1$, which in turn couples to two photons through a chargino loop (see Fig.~\ref{fig:diagram}). The dominant contribution to the thermally averaged cross-section for this process at zero temperature is given by \cite{Bergstrom:1997fh}  \begin{equation} \label{eq:sigma} \left<\sigma v\right>_{\gamma \gamma}=\frac{\alpha^2 m_{\chi_1^0}^2}{16 \pi^3}\left| \sum_{i=1,2} \frac{M_{\chi_i^{\pm}}m_{\chi_1^0}}{4m_{\chi_1^0}^2 \left(4m_{\chi_1^0}^2-m_{A_1}^2\right)} \hspace{2mm} g_{A_1 \chi_1^0} \hspace{2mm} g_{A_1 \chi_i^{\pm}} \hspace{2mm} F\left(\frac{m_{\chi_1^0}}{m_{A_1}},\frac{M_{\chi_i^{\pm}}}{m_{A_1}}\right)\right|^2 \end{equation} where the function $F(a,b)$ is defined by \begin{equation} F(a,b)\equiv \int_0^1 \frac{dx}{x}\log\left(\left|\frac{4ax^2-4ax+b}{b}\right|\right) \end{equation} and the couplings $g_{A_1 \chi_1^0}$, $g_{A_1 \chi_i^{\pm}}$ depend on the neutralino, chargino, and CP-odd Higgs diagonalizing matrices.  To compute these couplings, we use the Feynman rules found in Ref.~\cite{Ellwanger:2004xm}, appropriately modified to match our conventions for the neutralino and chargino matrices, which contain complex mass entries.  This cross-section is plotted as a function of $m_{A_1}$ for a particular choice of parameters, in Fig.~\ref{fig:cross_section}, which clearly shows the narrow resonant structure. 

%%%%%%%%%%%%%%%%%%%%%%%%%%%%%%%%%%%%%%%%%%%%%%%%%%%
%%%%%%%%%%%%%%%%%%%%%%%%%%%%%%%%%%%%%%%%%%%%%%%%%%%

\subsection{Suitable Higgs and Neutralino Sectors}\label{subsec:suitable}

Given our set-up, we can elucidate the parameter space regions capable of producing the gamma-ray line while satisfying all other dark matter and particle physics constraints.  As we show below, requiring a 130 GeV line from resonant neutralino annihilation restricts the NMSSM parameter space to a narrow region in which we can study electroweak baryogenesis and the electroweak phase transition, in addition to producing unambiguous predictions for several experimentally observable quantities, such as electric dipole moments and dark matter detection rates.

In general, the properties associated with the neutralino LSP depend sensitively on the details of the various parameters involved; this can be appreciated by considering the different benchmark points discussed in Refs.~\cite{Das:2012ys, Chalons:2012xf}.  For example, the annihilation cross-section into photons, Eq.~(\ref{eq:sigma}), is strongly affected by the mass splitting $\left|m_{A_1}-2m_{\chi_1^0}\right|$, as shown in Fig.~\ref{fig:cross_section}.  Correspondingly, other resonant processes, such as the $s$-channel neutralino pair annihilation into $b\bar{b}$ through $A_1$, also depend on the mass difference.  The details of the various resonant channels significantly affect both the zero-temperature and the finite-temperature annihilation cross sections (the latter being relevant for the calculation of the thermal relic density of dark matter). The amplitudes associated with these processes can however be tuned so that the neutralinos produce a 130 GeV gamma-ray line while satisfying all other indirect detection and relic density constraints, as we show here.  

Since we will be concerned with properties of the electroweak phase transition and baryogenesis which do not depend sensitively on the details of the resonance, it is sufficient, for our purposes, to consider the simple parameter choice $A_1= 2 m_{\chi_1^0}=260$ GeV and proceed to consider the implications for electroweak baryogenesis (a slightly off-resonance value would not at all affect the electroweak phase transition or the resulting baryon asymmetry).  From this starting point, we shall dial in the various parameters point-by-point to satisfy all of the phenomenological and observational constraints we describe below.

First and foremost, besides requiring the desired neutralino annihilation structure, demanding a 130 GeV LSP neutralino and the associated 260 GeV singlet-like $A_1$, we require a 125 GeV SM-like Higgs, in accordance with recent experimental findings from the LHC collaborations \cite{CMS, ATLAS}.  Given our parameter space, the requirements on the bino-like LSP and on $A_1$ lead us to vary $M_1$ and $A_{\lambda}$ in the range \begin{equation} \begin{aligned} &135 {\rm GeV} \hspace{.2cm} \leq \hspace{.2cm}  M_1 \hspace{.2cm} \leq \hspace{.2cm} 145 {\rm GeV} \\ &150 {\rm GeV}\hspace{.2cm}\leq \hspace{.2cm}A_{\lambda}\hspace{.2cm} \leq \hspace{.2cm}600 {\rm GeV}. \end{aligned} \end{equation} 
For each point in the $M_1$, $A_{\lambda}$ parameter space, we use the following strategy to choose values for the seven remaining parameters: 
\begin{enumerate} 
\item To obtain a Higgs mass of 125 GeV in the NMSSM without excessive tuning in the stop sector requires relatively large $\lambda$ and small $\tan \beta$, as seen from the tree-level inequality: \begin{equation}m_{h_1}^2\leq\left(\cos^22\beta+\frac{2\lambda^2\sin^22\beta}{g_1^2+g_2^2}\right)m_Z^2. \end{equation}  We take $\tan \beta$ in the range $1.7\leq \tan\beta\leq1.8$.  In principle $\lambda$ can be either positive or negative.  We focus on positive $\lambda$ and consider $0.6\leq \lambda\leq 0.8$ (see, e.g.\ Ref.~\cite{Chalons:2012xf} for a discussion of the case of $\lambda<0$).  For $\left|\lambda\right|$ much smaller than this value, one must rely heavily on stop loops to raise the Higgs mass.  Also, $\lambda$ determines the coupling of neutralinos to $A_1$, as well as the coupling of $A_1$ to photons, and so for much smaller $\left|\lambda\right|$ the neutralino annihilation cross-section into photons is suppressed.  For values $\lambda \gtrsim 0.7$, $\lambda$ becomes non-perturbative below the GUT scale; this can be remedied by including higher-dimension operators resulting from integrating out new physics which enters below the GUT scale\footnote{We will in fact assume that this is the case for our benchmark EWPT point which features $\lambda=0.75$.}.  

\item The pseudoscalar $A_1$ must be predominantly singlet-like to be compatible with indirect detection results.  The amount of mixing between $A_1$ and the MSSM-like CP-odd Higgs $A_2$ is governed by $\mathcal{M}_{P,12}$ in Eq.~(\ref{eq:mA}) and is minimized for \begin{equation} \label{eq:kap} \kappa \approx \frac{ \lambda A_{\lambda}}{2\mu}. \end{equation}  Given the relatively large values of $\lambda$ we consider, we take $\kappa \geq 0.3$.  For a given choice of $\kappa$, the $A_1-A_2$ mixing will vary point-by-point in the parameter space under consideration.  Therefore in some regions of parameter space the lightest pseudoscalar can obtain a large branching ratio into fermions and be incompatible with indirect detection constraints for a given mass difference $\left|m_{A_1}-2m_{\chi_1^0}\right|$.  As mentioned above (and discussed in more detail in Sec.~\ref{subsec:pheno}), one can typically dial in the details of the resonance to satisfy these constraints for a given point, however the BAU does not depend sensitively on this tuning.  

\item To obtain a lightest neutralino mass of 130 GeV, we must fix $\mu$ and $M_2$ or, equivalently, $\mu$ and $\Delta$ appropriately, where we define the quantity $\Delta$ via \begin{equation} M_2\equiv (\left|\mu\right|+\Delta)e^{i\phi}. \end{equation}  When considering CP-violation in Sec.~\ref{sec:EWB}, we will typically set the CP-violating phase $\phi$ to its maximal value, $\sin\phi=1$, in our calculations to show the maximum extent of the EWB parameter space, although viable regions will typically have phases of $\mathcal{O}(10^{-1})$.  In calculating the baryon asymmetry, $\Delta$ will govern the strength of the resonant CP-violating source. In considering the higgsino-gaugino CP-violating sources we will typically take $\Delta=0$ as an optimistic EWB scenario.  Given a particular choice of $\Delta$ and $\phi$, we fix $\mu$ by diagonalizing Eq.~(\ref{eq:neutralino}) and solving for $\mu$ such that $m_{\chi_1^0}=130$ GeV (note that we can rewrite $v_s=\mu/\lambda$).  This procedure fixes all the relevant parameters in the neutralino and chargino sectors.

 \item Finally, to obtain a large photon annihilation cross-section, we need the annihilation channel $\chi^0_1 \chi^0_1\rightarrow A_1$ to be near resonance at $T=0$, which implies $m_{A_1}\approx 260$ GeV.  As discussed above and shown in Fig.~\ref{fig:cross_section}, there is a narrow ($\lesssim 1$ GeV) window for which $\left<\sigma v\right>_{\gamma \gamma}$ is large enough to be compatible with the line.  Since the properties of the electroweak phase transition and baryogenesis are not sensitive to the precise value of $m_{A_1}$, we choose to sit exactly on top of the resonance, i.e. enforce $m_{A_1}=260$ GeV, by diagonalizing Eq.~(\ref{eq:mA}) and solving for the appropriate value of $A_{\kappa}$.  Therefore, at each point in the parameter space, $\left<\sigma_{\gamma \gamma}v\right>>10^{-27}$cm$^3/$s.  Once again, the precise mass splitting between $A_1$ and the LSP can typically be tuned point-by-point to produce the line while providing the correct relic density and satisfying the other indirect detection constraints as described below.
\end{enumerate}
The strategy outlined above is useful to automatically select the regions in the NMSSM producing the tentatively observed 130 GeV gamma-ray line, and provides an efficient way to study the properties of electroweak baryogenesis in these regions by exploring the remainder of the parameter space.  Note that we are not concerned with tuning or naturalness in this scenario, since we have narrowed in on this region by demanding consistency with the (tentative!) observation of a gamma-ray line which we postulate to be associated with dark matter pair annihilation. 

We shall now use our suitably selected Higgs and neutralino sectors to close in onto electroweak baryogenesis in regions of the NMSSM producing a 130 GeV line.  However, we first comment further on the impact of various other dark matter and particle physics constraints on the parameter space under consideration.

\subsection{Phenomenological Constraints}\label{subsec:pheno}

\begin{figure*}[!t]
\mbox{\hspace*{-1.2cm}\includegraphics[width=0.6\textwidth,clip]{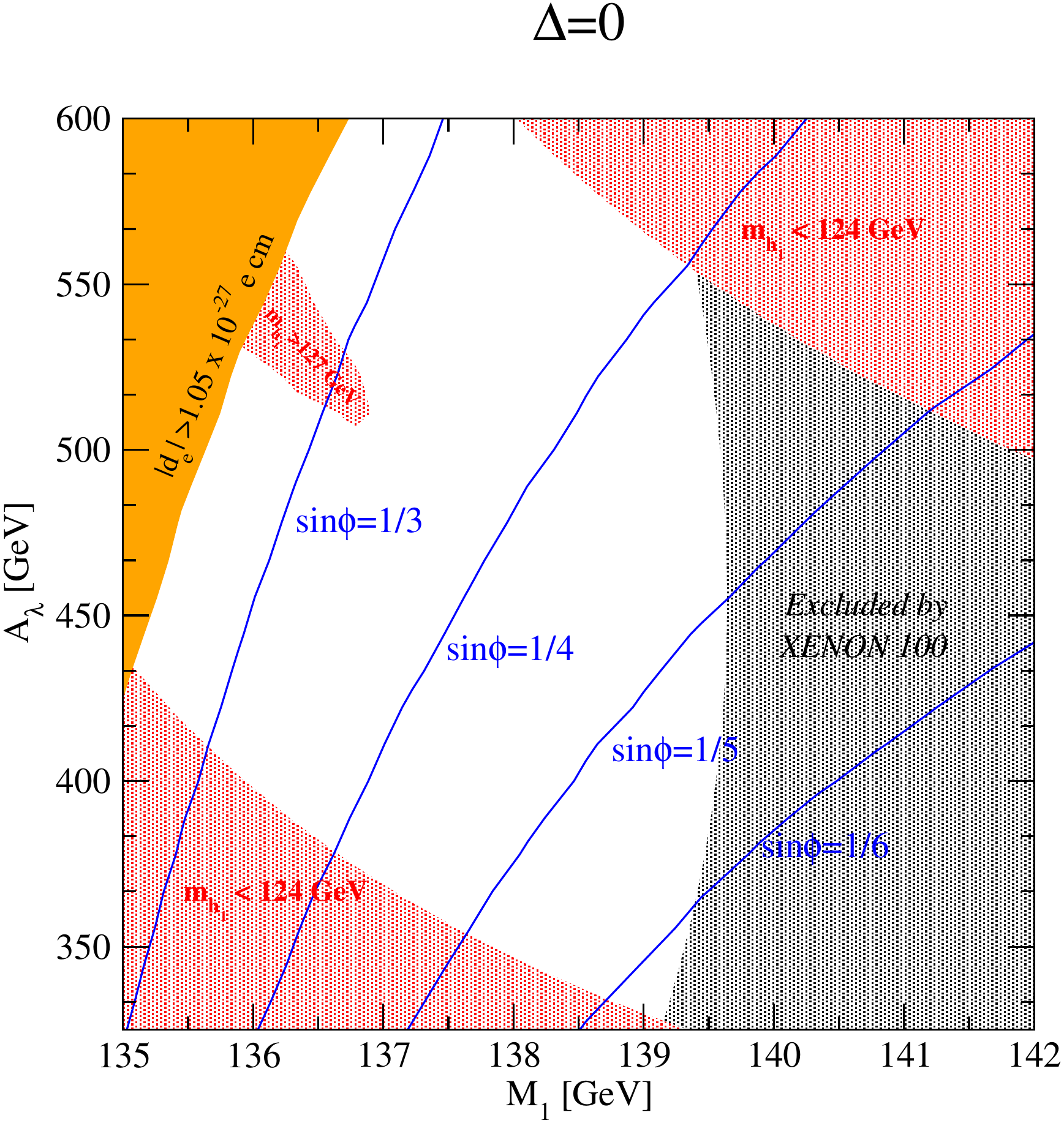}}\caption{\label{fig:results}\it\small  An example of the NMSSM parameter space for successful electroweak baryogenesis and a 130 GeV gamma-ray line.  Here we take $\lambda=0.6$, $\kappa=0.32$, $\tan\beta=1.8$ and $\Delta=0$ (so that the CP-violating sources are on resonance), while the rest of the parameters are chosen as described in Sec.~\ref{subsec:suitable} to be consistent with the Fermi line.   The gray shaded region is excluded by the XENON100 225 live day results, calculated with the default settings in \texttt{MicrOmegas}.  Red shaded regions are excluded by measurements of the Higgs mass (although these regions can be shifted around by changing e.g.\ the squark masses).  The orange shaded region is excluded by the non-observation of an electric dipole moment of the electron.  The blue contours correspond to points consistent with the observed baryon-to-entropy ratio of the universe for different values of the CP-violating phase $\phi$. }
\end{figure*}

The NMSSM parameter space of interest features relatively light neutralino, chargino, and Higgs sectors and is thus quite constrained on multiple fronts.  Here we highlight the most important constraints on the parameter space and consider their impact on our current set-up.  We use \texttt{NMSSMTools 3.2.1}\cite{NMSSMTools} and \texttt{MicrOmegas 2.4.5}\cite{micromegas} to calculate the various cross-sections and quantities of interest.  We summarize in Fig.~\ref{fig:results} the impact of the constraints we consider here (and that we discuss in detail below) on the relevant parameter space, for the particular choice $\lambda=0.6$, $\kappa=0.32$, and $\tan\beta=1.8$ as an illustrative example.  In these calculations, we take $M_2$ to be real; since the LSP has only a very small wino component across the parameter space, and since the other neutralinos and charginos are significantly heavier than the lightest neutralino, the DM constraints will be largely unaffected by allowing $M_2$ to be complex.  The Higgs couplings are also insensitive to $\phi$. 

%%%%%%%%%%%%%%%%%%%%%%%%%%%%%%%%%%%%%%%%%%%%%%%%%%%
%%%%%%%%%%%%%%%%%%%%%%%%%%%%%%%%%%%%%%%%%%%%%%%%%%%

\subsubsection {Indirect Dark Matter Detection and Thermal Relic Density}\label{subsec:indirect}

Indirect detection places important constraints on the parameter space in question.  In considering $m_{A_1}\approx 2m_{\chi_1^0}$, there will also be a resonant tree-level neutralino annihilation channel into quark-antiquark, and especially $b \bar{b}$, final states, eventually leading to gamma rays via hadronization producing neutral ions. The lack of an excess of gamma-rays associated with this emission puts constraints on the branching ratio for neutralino pair-annihilation into, e.g., $b\bar b$  \cite{Fermi}.  As mentioned above, however, one can generally dial in the mass splitting $\left|m_{A_1}-2m_{\chi_1^0}\right|$ to obtain both $\langle\sigma_{\gamma \gamma}v\rangle \gtrsim 10^{-27}$ cm$^3/$s and  $\langle\sigma_{b \bar{b}}v\rangle \lesssim 10^{-24}$ cm$^3/$s as required by Fermi observations \cite{Fermi} of the diffuse gamma ray background  (see e.g.\ the benchmark point in Table~\ref{table1}).  Additionally, neutralino annihilation into $W^+W^-$ will receive a contribution at tree-level from the pseudoscalar channel; however, this contribution  also typically falls well beneath the $10^{-24}$ cm$^3/$s bound from Fermi by adjusting $m_{A_1}$.  Consequently, this tuning allows one to satisfy all continuum gamma-ray constraints \cite{Cohen:2012me} while reproducing the observed intensity of the 130 GeV line, something that cannot be done in the MSSM.  The parameter space we consider for electroweak baryogenesis can thus be dialed in to agree with indirect detection results without drastically affecting the details of the electroweak phase transition or the generation of the baryon asymmetry.

Similar reasoning applies to the DM thermal relic abundance.  For $\chi_{1}^0$ to be a suitable thermally-produced dark matter candidate, it must be compatible with the bounds on the relic density from WMAP7 \cite{WMAP}: $\Omega_{DM} h^2 =0.112\pm .011$.  While at zero-temperature the neutralino sits very close to the pseudoscalar resonance, at the freeze-out temperature $T_{\rm f.o.}\sim m_{\chi_{1}^0}/20\approx 6.5$ GeV, the resonance is shifted higher by about $10$ GeV for the case of $m_{A_1}=260$ GeV.  This can be seen by evaluating the thermally-averaged center-of-mass (C.O.M.) energy, $\langle s\rangle$, at $T=T_{\rm f.o.}$, given by \begin{equation}   \langle s \rangle \simeq 4 m_{\chi_1^0}^2+6 m_{\chi_1^0} T_{\rm f.o.} \simeq 270 \hspace{2mm} {\rm GeV}. \end{equation}  However, in evaluating $\langle \sigma v\rangle$ at $T_{\rm f.o.}$, one integrates over center-of-mass energies, and hence effectively picks up contributions from the resonances, which decrease as one moves $\langle s \rangle$ further away from $4 m_{\chi_1^0}^2$.   Therefore, as is the case for the zero-temperature cross-sections, by dialing in the detailed neutralino and pseudoscalar masses, as well as the $A_1-A_2$ mixing, one can typically achieve a total annihilation thermally averaged cross-section of $\langle \sigma v \rangle \sim 3\times 10^{-26}$ cm$^3/$s required to obtain the correct relic density.

Previous studies \cite{Das:2012ys, Chalons:2012xf} have relied on a sizable higgsino component in the LSP to drive the relic density down.  However, this requires small values of $\mu$ which are difficult to reconcile with the most recent direct detection constraints, except in the case of cancellations which can occur for negative $\mu$ as exploited in Ref.~\cite{Chalons:2012xf} (we have found it difficult to achieve a strongly first-order EWPT consistent with the 130 GeV line for the $\mu<0$ case, but it may still be possible).  Another possibility is to open a co-annihilation channel by e.g.\ allowing a light stau \footnote{Of course with CP-violation in the gaugino sector one must verify that such a light slepton satisfies constraints from EDMs.} with mass near 130 GeV to drive the relic density down.  Light staus are not yet significantly constrained by LHC searches and, interestingly, they could provide an explanation of the enhanced Higgs diphoton rate as observed by ATLAS, albeit for large $\tan \beta$ (see e.g.\ Ref.~\cite{Carena:2012gp}).  We do not pursue these avenues further, but emphasize that we find that the relic density (and the zero-temperature neutralino annihilation cross-sections) can be made to agree with observations in this scenario by tuning or other mechanisms that do not significantly affect the properties of the EWPT nor the calculation of the baryon asymmetry.  Consequently, we do not focus on the detailed bounds from indirect detection or the thermal relic abundance point-by-point in our present study of EWB in this scenario, but we do emphasize that these constraints can all be met in principle, as illustrated by a worked-out example in the EWPT benchmark point we show explicitly in Table~\ref{table1}.

\subsubsection{Direct Detection}

Unlike the case of indirect detection and relic density constraints, the bounds from DM direct detection (i.e. the scattering of the lightest neutralino off of nucleons) do not depend sensitively on the details of the resonance, but rather on the composition of the lightest neutralino.  This in turn depends on $M_1$: larger values of $M_1$ require smaller values of $\mu$ to obtain $m_{\chi_1^0}=130$ GeV and consequently enhance the spin-independent neutralino-proton cross section.  

We require that the LSP satisfy the current upper bound from XENON100 for a 130 GeV WIMP for the spin-independent cross-section\footnote{We also consider the bound on the spin-dependent cross-section, but the corresponding constraints are much weaker than those on $\sigma_{SI}$ in our scenario}, $\sigma_{SI}\lesssim 3\times 10^{-9}$ pb \cite{xenon100}.  We show the impact of this constraint on our parameter space in Fig.~\ref{fig:results}: points excluded by XENON100 are shown in the gray shaded region.  These bounds are computed assuming default values for the various underlying parameters, such as the quark content of the nucleon, local distribution of dark matter, etc. We employ the \texttt{MicrOmegas 2.4.5}\cite{micromegas} package for the calculation of the relevant scattering cross section, and employ the default parameters thereof. As expected, points with smaller $\mu$ values, and hence a larger higgsino component in $\chi_1^0$, are ruled out.  

We note here that the exclusions are somewhat stronger than those reported in Ref.~\cite{Das:2012ys} due to the release of the 2012 XENON results (and consequently the window for $m_{A_1}$ is somewhat more constrained than that in Ref.~\cite{Das:2012ys}).  Since these limits depend on parameters affected by significant uncertainty, they should also be taken with a grain of salt.  For example, by considering the strange quark content of the nucleons near the end of the error bars from Ref.~\cite{Thomas:2012tg} ($\sigma_{\pi N}=39$ MeV, $\sigma_0=43$ MeV), one can push the XENON limits out to allow $M_1$ up to $\sim 145$ GeV consistent with the 2012 XENON100 results (see e.g.\ the EWPT benchmark point in Table~\ref{table1}).

\subsubsection{Higgs Constraints}

The lightest CP-even Higgs in our scenario is SM-like.  We require that $124$ GeV $<m_{h_1}<127$ GeV, in agreement with results from ATLAS \cite{ATLAS} and CMS \cite{CMS}.  The region of parameter space incompatible with these results is shown in Fig.~\ref{fig:results} by points within the red shaded regions.  We have also checked against constraints from $h_1\rightarrow b\bar{b}$, $\tau \tau$, etc. as implemented in \texttt{NMSSMTools 3.2.1}\cite{NMSSMTools}.  The couplings of $h_1$ to the various SM fermions and gauge bosons all fall within $\sim 3\%$ of the corresponding SM predictions, hence well within experimental limits.

The lightest CP-odd Higgs must also be compatible with collider searches.  In particular, we verified that the couplings of $A_1$ to $b\bar{b},$ $\tau\tau$ are small compared to that of the SM-like Higgs for compatibility with LHC results.  In the parameter space under consideration, we find that the couplings of $A_1$ are at most of order $1\%$ of the SM Higgs couplings.

\subsubsection{Other Considerations}

There are several other constraints which are in fact satisfied over nearly all of the parameter space we consider.  Constraints from LEP on light charginos are everywhere satisfied, since charginos are always heavier than the 130 GeV LSP.  Also, constraints from $B$-physics, as implemented in \texttt{NMSSMTools 3.2.1}, do not constrain the parameter space since we consider small values of $\tan\beta$.  Finally, we have also verified the absence of unphysical global minima of the effective potential for all points we consider, as well as the absence of Landau poles below the GUT scale, with the exception of the EWPT benchmark point, for which we take $\lambda=0.75$.  As discussed above, this issue can be remedied with the modest assumption of new physics entering below the GUT scale.

In summary, Fig.~\ref{fig:results} shows that there exist regions of NMSSM parameter space consistent with a 130 GeV gamma-ray line, a 125 GeV SM-like Higgs, and which can satisfy all relevant dark matter and experimental particle physics constraints.  We can now proceed to investigate the phenomenology and properties of electroweak baryogenesis in these regions.

\section{The Electroweak Phase Transition} \label{sec:EWPT}

Successful electroweak baryogenesis requires a strongly first-order electroweak phase transition.  In the absence of a strongly first-order transition, $SU(2)$ sphaleron processes, which provide the necessary baryon number violation, are unsuppressed in the broken electroweak phase and tend to wash out any existing generated baryon asymmetry.  The strength of the phase transition can be parametrized by the order parameter $\varphi(T_c)/T_c$, where $T_c$ is the critical temperature, defined as the temperature for which the symmetric and broken phases are degenerate\footnote{Note that this quantity is not gauge invariant, see e.g.\ the discussion in Ref.~\cite{Patel:2011th, Wainwright:2012zn}.}.  To prevent sphaleron washout requires $\varphi(T_c)/T_c\gtrsim 1$, which we take as the definition of a ``strongly first-order" transition\footnote{
More precisely, one should actually consider the system at the nucleation temperature, $T_n$.  However, the amount of supercooling in this model is small, and for simplicity we assume that $T_n \approx T_c$ as in previous work.
}.  As we will show in this section, this requirement can be readily satisfied in the region of the NMSSM compatible with the 130 GeV gamma-ray line and without relying on a light stop squark, as is instead typically required in the MSSM \cite{Balazs:2004bu, Carena:2008vj}. 
  
The strength of the electroweak phase transition is governed by the finite-temperature effective potential, which comprises several parts: the tree-level scalar potential, zero-temperature quantum corrections, finite-temperature quantum corrections, and thermal mass terms. The tree-level potential comes directly from the superpotential (Eq.~(\ref{eq:super})) and the soft supersymmetry-breaking terms (Eq.~(\ref{eq:soft})):
\begin{multline}
V_0(h_u, h_d, s) =
\frac{1}{32}(g_1^2+g_2^2) \left( h_u^2 - h_d^2\right)^2 + \frac{1}{4} \kappa^2 s^{4} - \frac{1}{2} \lambda \kappa s^2 h_u h_d + \frac{1}{4} \lambda^2 \left(h_d^2 h_u^2 + s^2 \left(h_d^2 + h_u^2\right)\right) \\
 + \frac{\sqrt{2}}{6}  \kappa A_{\kappa}  s^3 - \frac{\sqrt{2}}{2} \lambda A_{\lambda}   s h_u h_d 
+ \frac{1}{2} m_d^2 h_d^2  + \frac{1}{2} m_u^2 h_u^2 + \frac{1}{2} m_s^2 s^2.
\end{multline}
The fields $h_u$, $h_d$, and $s$ are defined by
\begin{align}
H_u = \frac{1}{\sqrt{2}} \begin{pmatrix} 0\\h_u\end{pmatrix};\;
H_d = \frac{1}{\sqrt{2}} \begin{pmatrix} h_d\\0\end{pmatrix};\;
S = \frac{1}{\sqrt{2}} s.
\end{align}
We assume that the scalar fields are real at all temperatures, and we do not consider charged vacua (although we do ensure that the potential is stable in the charged and imaginary directions).

Using $\overline{MS}$ renormalization, the one-loop zero-temperature quantum corrections are
\begin{align}
V_1(T\!=\!0) = \sum_i \frac{\pm n_i}{64\pi^2} m_i^4 \left[\log\left(\frac{m_i^2}{\Lambda^2}\right) -c\right],
\end{align}
where $m_i^2$ are the (possibly negative) field-dependent mass-squared values, $n_i$ are their associated number of degrees of freedom, $\Lambda$ is the renormalization scale, and $c=\tfrac{1}{2}$ for the transverse polarizations of gauge bosons while $c=\tfrac{3}{2}$ for their longitudinal polarizations and for all other particles. The plus and minus signs are for bosons and fermions, respectively. The sum over the relevant particles $i$ include all standard model particles (although we ignore fermions lighter than the bottom quark), the physical Higgs and other scalar particles, their associated Goldstone bosons, the neutralinos and the charginos. We work in Landau gauge where the ghost bosons decouple and need not be included in the spectrum. The one-loop potential contains explicit gauge-dependence which cancels with the implicit gauge-dependence of the vevs at every order in $\hbar$ (for recent discussions of gauge dependence in effective potentials, see e.g.\ Refs.~\cite{Patel:2011th, Wainwright:2011qy, Wainwright:2012zn, Garny:2012cg}). As is common practice, we do not consider the effects of the implicit gauge-dependence, and therefore our results will contain gauge artifacts. However, our primary purpose in examining the effective potential is to estimate whether or not a first-order phase transition is possible, and for this purpose a rough calculation with gauge-dependence is acceptable.

We calculate the neutralino masses from Eq.~(\ref{eq:neutralino}) above. The scalar mass matrix is given by taking the second derivative of the tree-level potential, but including CP-odd and charged directions. This yields a block-diagonal $10\times10$ matrix, with blocks consisting of CP-even states (3 degrees of freedom), CP-odd states (3 degrees of freedom), and two blocks of charged Higgses (4 degrees of freedom) (see Appendix~\ref{sec:scalar_masses} for details).

The finite-temperature contributions are
\begin{align}
V_1(T\!>\!0) = V_1(T\!=\!0) + \frac{T^2}{2\pi^2} \sum_i n_i J_\pm \left(\frac{m_i^2}{T^2}\right),
\end{align}
where
\begin{align}
J_\pm(x^2) \equiv \pm \int_0^\infty dy \; y^2 \log\left(1 \mp e^{-\sqrt{y^2+x^2}}\right)
\end{align}
and again the upper (lower) signs correspond to bosons (fermions). At high temperature, the validity of the perturbative expansion of the effective potential breaks down. Quadratically divergent contributions from non-zero Matsubara modes must be re-summed through inclusion of thermal masses in the one-loop propagators \cite{gross1981,parwani1992}. This amounts to adding thermal masses to the longitudinal gauge boson degrees of freedom and to all of the scalars (see Appendix~\ref{sec:scalar_masses}).

The full one-loop effective potential is
\begin{align}
V(h_u, h_d, s, T) = V_0(h_u, h_d, s) + V_1(T\!=\!0) + \frac{T^2}{2\pi^2} \sum_i n_i J_\pm \left(\frac{m_i^2}{T^2}\right)
\end{align}
where the masses $m_i^2$ are field-dependent and include thermal mass corrections.

The important qualitative feature of the finite-temperature contribution is that it lowers the effective potential anywhere $m_i^2/T^2$ is small. To get a strongly first-order phase transition, we need to sharply lower the potential near the symmetric phase without significantly lowering it in the broken phase so that the two phases may be degenerate with a sizable barrier. Therefore, a strongly first-order transition demands either numerous heavy field-dependent particles (such that they are massless in the symmetric phase and heavy in the broken phase), or a tree-level contribution to the barrier separating the two phases. In the standard model, the electroweak phase transition is not strongly first-order. There are no heavy bosons (relative to the Higgs, which sets the relevant scale), and at high temperature the contribution of heavy fermions (top quarks) does not increase the barrier since $J_-(x^2)$ does not contain any cubic terms. 

The particle spectrum in the NMSSM may seem somewhat promising, since there are additional heavy masses in the Higgs sector and field-dependent neutralino masses, but these are not enough to guarantee a strong transition. Since many more particles couple to the Higgs than to the singlet, finite-temperature effects drive $\langle h_u \rangle$ and $\langle h_d \rangle$ to zero at temperatures well below the point at which they drive $\langle s \rangle$ to zero. Therefore, $s$ can be large on either side of electroweak symmetry breaking, and some of the new particle masses that depend on $s$ can be heavy even in the symmetric phase.

However, the NMSSM can succeed in producing a strongly first-order transition through its tree-level contributions. If the transition occurs both in the Higgs and singlet directions simultaneously, and if the singlet vev is non-zero in the electroweak symmetric phase just above the transition, then terms like $s^2 h^2$ and $s h^2$ both contribute effective cubic terms to the potential which can increase the barrier between the the symmetric and broken phases.

We calculate the phase transition using the software package {\tt CosmoTransitions} \cite{Wainwright:2011kj}. We input the above definition of the effective potential, find the necessary soft-breaking masses that produce desired values for $\tan\beta$ and $\mu$ via a minimization procedure, and choose a renormalization scale $\Lambda$ such that the one-loop minimum does not drastically differ from its tree-level value. This last point requires a certain amount of finesse since the top-quark contribution to the zero-temperature one-loop potential tends to be fairly large. The {\tt CosmoTransitions} package traces the broken electroweak phase up in temperature until it disappears, and then traces the symmetric phase down and checks for an overlap. If there is one, it calculates the temperature of degeneracy (the critical temperature) and the separation between the phases. If there is no overlap, then the transition is necessarily second-order.

\begin{table}[tc]
\centering

 \begin{tabular}{l c || l c}
\hline \hline
$\lambda$ \hspace{1cm} & 0.75 & $m_{A_1}$ [GeV] \hspace{1cm} & 261.26\\
$\kappa$ \hspace{1cm}  & 0.45 & $m_{\chi_1^0}$ [GeV] \hspace{1cm}  & 130.72 \\
$\tan \beta$ \hspace{1cm}  & 1.7 & $\langle \sigma v \rangle_{b\bar{b}}$ [$cm^3/s$] \hspace{1cm}  & $3.07  \times10^{-26}$ \\
$A_{\lambda}$ [GeV] \hspace{1cm}  &  545.0 &  $\langle \sigma v \rangle_{\gamma\gamma}$ [$cm^3/s$] \hspace{1cm}  & $1.54  \times10^{-27}$   \\
$A_{\kappa}$ [GeV] \hspace{1cm} & -88. 0&$\sigma^{\rm SI}_P$ [pb] \hspace{1cm}& $2.8 \times 10^{-9}$ \\
$\mu$ [GeV] \hspace{1cm} &275.8 & $\sigma^{\rm SD}_P$ [pb] \hspace{1cm}& $1.4 \times 10^{-6}$ \\
$M_1$ [GeV] \hspace{1cm}  & 143.5& \underline{EWPT Properties:}&\\
$M_2$ [GeV]\hspace{1cm}  & 635.5& $T_{c}$ [GeV]  &72.3 \\
$m_{h_1}$ [GeV] \hspace{1cm} & 126.4 & $\varphi(T_c)/T_c$&1.14\\

\hline \hline

\end{tabular}
\caption{\label{table1} \it\small Benchmark Point in the NMSSM with a strongly first-order EWPT and a 130 GeV line. We use a renormalization scale of $\Lambda=100$ GeV in the effective potential.}
\end{table}

The region of the NMSSM consistent with the 130 GeV Fermi line can in fact accommodate a strongly first-order phase transition.  The barrier has large tree-level contributions and in particular does not require an additional light scalar. As a proof of principle, we outline a benchmark point consistent with a 125 GeV Higgs, 130 GeV Fermi line, and a strongly first-order electroweak phase transition in Table~\ref{table1}.  This point has an EWPT at $T_c= 72.3$ GeV with order parameter $\varphi(T_c)/T_c=1.14$ and is consistent with all other relevant phenomenological constraints \footnote{As mentioned previously, we can invoke some higher-dimension operators to render $\lambda$ perturbative below the GUT scale.}.  The spin-dependent and --independent neutralino-proton scattering cross-section for the point in Table~\ref{table1} is computed taking $\sigma_{\pi N}=39$ MeV, $\sigma_0=43$ MeV for the strange quark content of the proton and is thus rather optimistic.  Also, note that we do not show the relic density for the specified point.  Since we are near a resonance, as discussed in Sec.~\ref{subsec:pheno}, the relic density calculation should be performed to loop level -- something which is not implemented in \texttt{MicrOmegas} \footnote{We have also found a suspected numerical issue with the \texttt{MicrOmegas 2.4.5} calculation of the relic density near the resonance.  There is a very sharp increase in the annihilation cross section right above $m_{A_1}=2 m_{\chi_1^0}$ which we believe is unphysical.  Since the zero-temperature total-annihilation cross section is of order $\langle \sigma v\rangle\sim 10^{-26}$ cm$^3/$s, by the arguments in Sec.~\ref{subsec:indirect} the thermally averaged cross-section at freeze-out should be smaller than this since the resonance is effectively shifted.  Instead, we find a drop of four orders of magnitude in the relic density which is quite suspect.}.  Neglecting one-loop processes, the relic density for this point may be too large.  We have checked, however, that at tree-level and neglecting the contribution from the resonances, one can introduce a light slepton with $M_{R_3}\sim 140$ GeV which will set $\Omega h^2 =0.11$ for the parameters shown.  Since $\tan \beta$ is small, the presence of such a light slepton will not affect the properties of the EWPT.  Thus, we are confident that a proper one-loop calculation of the relic density for the benchmark point in Table~\ref{table1} will yield a relic density compatible with observation, albeit with some possible changes to the parameters or the introduction of a co-annihilation channel which will not substantially affect the EWPT.  

\begin{figure*}[t]
\mbox{\hspace*{0cm}\includegraphics[width=0.6\textwidth,clip]{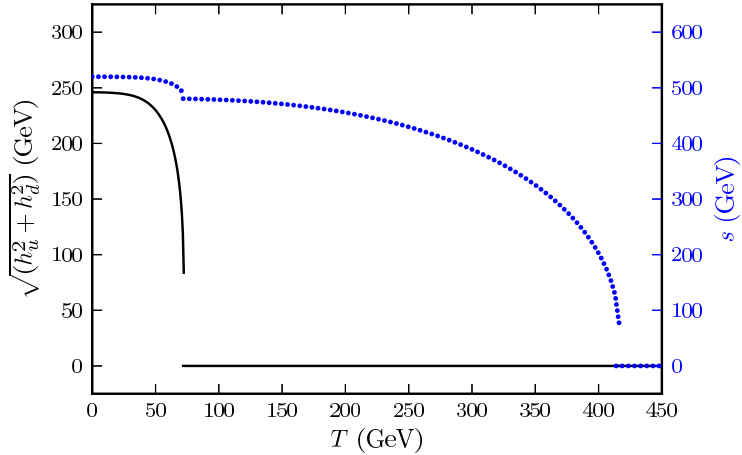}}\caption{\label{fig:phase_structure}\it\small The phase structure for the benchmark point with first-order phase transitions. The dotted line gives the temperature-dependent singlet field values, and the solid line gives the temperature-dependent Higgs doublet field values.}
\end{figure*}

Fig.~\ref{fig:phase_structure} shows the field evolution as a function of temperature for the benchmark point in Table~\ref{table1}. This makes the location of the phase transitions obvious: first-order phase transitions can happen anywhere there is a discontinuous jump in the vacuum expectation values. A second-order transition, if there were one, would be distinguished by a continuous line of vacuum expectation values with discontinuous first derivatives.

\begin{figure*}[t]
\mbox{\hspace*{0cm}\includegraphics[width=0.6\textwidth,clip]{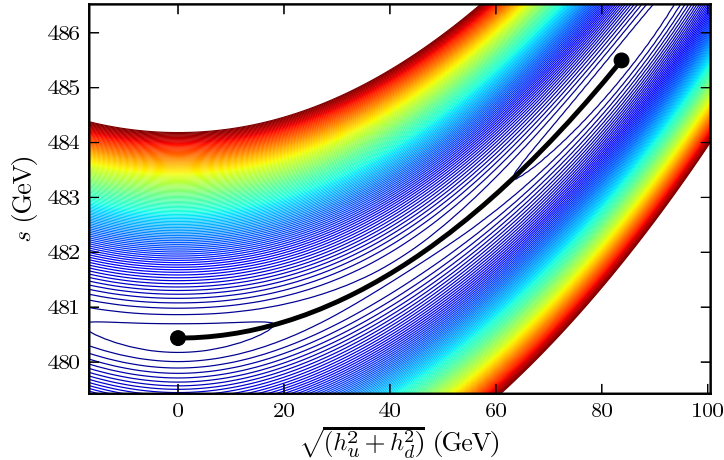}}\caption{\label{fig:tunneling}\it\small A contour plot of the effective potential just below the critical temperature. The electroweak broken minimum is represented by the dot on the upper-right, while the symmetric minimum is on the lower left. The actual tunneling happens along the curved solid black line. }
\end{figure*}

Fig.~\ref{fig:tunneling} shows the field configuration at the critical temperature of electroweak symmetry breaking. All three fields --- $s$, $h_u$ and $h_d$ --- change values when tunneling from the high-temperature to the low-temperature minimum. We calculate the tunneling direction (denoted by a thick black line) using the {\tt CosmoTransitions} package, where by ``tunneling direction'' we mean the path through field space that one would travel when crossing a bubble wall. The path is curved in the $s-h_u$ and $s-h_d$ planes, but is approximately straight in the $h_u-h_d$ plane ($\Delta\beta \ll 1$).

While we did not perform a systematic study of the NMSSM parameter space compatible with a strongly first-order transition (see e.g.\ Refs.~\cite{ Funakubo:2005pu, Pietroni:1992in} for previous work in this direction), there are some common traits between the viable points we have found.  Restricting ourselves to the case of positive $\lambda$, $\kappa$, $\mu$, and $A_{\lambda}$, we find that a strongly first-order phase transition typically requires $ \lambda \gtrsim 0.6$, $\kappa \lesssim 0.6$, $A_{\lambda} \gtrsim 500$ GeV, and $\mu\lesssim 350$ GeV.  This seems to be consistent with our intuition: increasing the strength of the cubic terms in the effective potential and decreasing the singlet vev tends to strengthen the transition.  Note that, for all the points we considered, the transition tends to happen in two steps: the system transitions away from $\langle s \rangle =0$ at a high temperature, around 300--400 GeV; while electroweak symmetry breaking happens much later, at a temperature around or below 100 GeV.

\section{Computing the Baryon Asymmetry} \label{sec:EWB}

The discussion in the previous section makes it clear that a strongly first-order EWPT can occur in the NMSSM region of parameter space compatible with the Fermi 130 GeV line.  We now turn our attention to the CP-violating sources also required for electroweak baryogenesis, and to the detailed requirement of producing the correct amount of baryon asymmetry in the early universe, parametrized by the baryon-to-entropy ratio\footnote{For concreteness and consistency with previous studies, we take $Y_B=9.1\times 10^{-11}$}, $Y_B\sim 10^{-10}$.  As we show in this section, CP-violating higgsino-gaugino sources can be very efficient in the NMSSM regions of interest and potentially source the observed baryon asymmetry of the universe.

In electroweak baryogenesis, the baryon asymmetry is produced by $SU(2)$ sphalerons acting on a net left-handed chiral density, $n_L$.  To determine $n_L$, we must solve a set of quantum transport equations for each of the relevant particle densities contributing to the LH charge density.  For each of these charge densities, $n_i$, the Schwinger-Dyson equations yield the continuity equations \cite{Lee:2004we} \begin{equation} \label{eq:cont} \frac{\partial n_i}{\partial x_0}+\nabla \cdot \mathbf{}j_i(x)=S_i(x).\end{equation}  The RHS of the above equation contains both CP-conserving and CP-violating contributions.  For the case of Dirac fermions, the sources are given by \begin{equation}\label{eq:source1} \begin{aligned} S_i(x)=\int d^3 z \int_{-\infty}^{x_0}dz_0 &\operatorname{Tr}\left[ \Sigma^{>}(x,z)G^<(z,x)-G^>(x,z)\Sigma^<(z,x)\right.\\&  \left.+G^<(x,z)\Sigma^>(z,x)-\Sigma^<(x,z)G^>(z,x)\right] \end{aligned} \end{equation} where $G^{<,>}$, $\Sigma^{<,>}$ are Green's functions and self-energies, respectively, in the closed time path formalism (see e.g.\ Ref.~\cite{Lee:2004we} for details).  We focus here on the case of gaugino-higgsino sources, and compute the quantities $S_{\widetilde{H}^{0,\pm}}$ in the Higgs vev-insertion approximation, which we describe in more detail below (see e.g.\ Ref.~\cite{Kozaczuk:2012xv} for a recent discussion on scalar sources in the MSSM).

\subsection{The VEV-Insertion Approximation}

The CP-violating interactions we consider involve the scattering of higgsinos and gauginos with the spacetime-dependent Higgs vevs in the bubble wall.  In what follows we parallel the derivations for the corresponding quantities in the MSSM  found in Ref.~\cite{Lee:2004we}.  We will assume that the necessary CP-violating phase $\phi$ is that of the wino soft SUSY-breaking mass $M_2$ (in fact, the relevant phase is the relative phase between $M_{1,2}$ and $\mu$, however as discussed previously we take $\mu$, $M_1$ to be real to avoid large spontaneous CP-violating effects in the computation of the various dark matter properties). The part of the NMSSM Lagrangian giving rise to the relevant CP-violating interactions is then given, in terms of four-component spinors, by: \begin{equation} \label{eq:Lagrangian} \mathcal{L}_{\rm int} \supset -\frac{g_2}{\sqrt{2}} \bar{\Psi}_{\widetilde{H}^0}\left[v_d(x) P_L+e^{i \phi}v_u(x)P_R\right]\Psi_{\widetilde{W}^0}-g_2 \bar{\Psi}_{\widetilde{H}^+}\left[v_d(x)P_L+e^{i \phi}v_u(x)P_R\right] \Psi_{\widetilde{W}^+} + h.c. \end{equation} where $P_{L,R}$ are the usual projection operators.

The spinors $\Psi_{\widetilde{H}^{0,\pm}}$ satisfy Dirac equations with a spacetime-varying mass $\mu(x)$.  As discussed in Sec.~\ref{sec:EWPT}, the profile $\mu(x)$ depends on the detailed properties of the phase transition at each point in parameter space.  In the region of interest, however, the singlet vev does not change very significantly during the EWPT.  
Consequently, even though the variation of the singlet vev was crucial for achieving a strongly first-order phase transition, we ignore its space-dependence here\footnote{The spacetime-dependence of $\mu$ can introduce novel sources of CP-violation in the NMSSM; see e.g.\ Ref.~\cite{Huber:2006wf}} and approximate $\mu(x)$ by its value after the EWPT, $\mu(x)\simeq \mu$.
Then the mode expansions for the operators in the Lagrangian Eq.~(\ref{eq:Lagrangian}) are the same as in the MSSM case and so the resulting source from Eq.~(\ref{eq:source1}) matches that of the MSSM in the vev-insertion approximation: \begin{equation} \begin{aligned} \label{eq:source2} S_{\widetilde{H}^{\pm}}(x)&=\int d^4z  \sum_{j=A,B}\left\{\left[g_j(x,z)+g_j(z,x)\right]\operatorname{Re} \operatorname{Tr}\left[G_{\widetilde{W}^{\pm}}^>(x,z) G_{\widetilde{H}^{\pm}}^<(z,x)-G_{\widetilde{W}^{\pm}}^<(x,z)G_{\widetilde{H}^{\pm}}^>(z,x)\right]_j \right.\\ & \left. +i\left[g_j(x,z)-g_j(z,x)\right] \operatorname{Im}\operatorname{Tr}\left[G_{\widetilde{W}^{\pm}}^>(x,z)G_{\widetilde{H}^{\pm}}^<(z,x)-G_{\widetilde{W}^{\pm}}^<(x,z)G_{\widetilde{H}^{\pm}}^>(z,x)\right]_j\right\} \end{aligned}\end{equation} where the sum over $A$, $B$ is over contributions arising from momentum and mass terms in the spectral function, respectively, and where \begin{align} g_A(x,y)&\equiv \frac{g_2^2}{2}\left[v_d(x)v_d(y)+v_u(x)v_u(y)\right] \\ g_B(x,y)&\equiv \frac{g_2^2}{2}\left[v_d(x)e^{-i\phi}v_u(y)+e^{i\phi}v_u(x)v_d(y)\right]. \end{align}  

The rest of the derivation proceeds as in the MSSM case, i.e.\ by performing a derivative expansion in $g_{A,B}(x,z)$ around $z=x$.  The CP-conserving sources arise from the terms in Eq.~(\ref{eq:source2}) symmetric under the interchange of $x\leftrightarrow z$ and so appear at zeroth order in this expansion, while the CP-violating sources arise at first-order.  In particular, performing the integration for the CP-violating contribution yields \begin{equation} \label{eq:source3} \begin{aligned} S^{\slashed{\rm CP}}_{\widetilde{H}^{\pm}}= \frac{g_2^2}{\pi^2}v(x)^2&\dot{\beta}(x) M_2 \mu \sin\phi \int_0^{\infty}\frac{dk k^2}{\omega_{\widetilde{H}}\omega_{\widetilde{W}}}\operatorname{Im}\left\{ \frac{n_F(\mathcal{E}_{\widetilde{W}})-n_F(\mathcal{E}^*_{\widetilde{H}})}{(\mathcal{E}_{\widetilde{W}}-\mathcal{E}^*_{\widetilde{H}})^2}-\frac{n_F(\mathcal{E}_{\widetilde{W}})+n_F(\mathcal{E}_{\widetilde{H}})}{(\mathcal{E}_{\widetilde{W}}+\mathcal{E}_{\widetilde{H}})^2}\right\} \end{aligned} \end{equation} where $\omega^2_{\widetilde{H},\widetilde{W}}\equiv \left| \bf{k}\right|^2+M^2_{\widetilde{H},\widetilde{W}}$ (the masses here include thermal contributions, $\delta_{\widetilde{H},\widetilde{W}}$), $\mathcal{E}_{\widetilde{H},\widetilde{W}}\equiv \omega_{\widetilde{H},\widetilde{W}} - i\Gamma_{\widetilde{H},\widetilde{W}}$ (here the $\Gamma_{\widetilde{H},\widetilde{W}}$ are the thermal widths of the higgsinos and winos in the plasma), and $n_F$ is the Fermi distribution function. The corresponding expressions for the CP-conserving (and neutral higgsino CP-violating) sources can be found in Ref.~\cite{Lee:2004we} with the appropriate replacements.

The CP-violating source in Eq.~(\ref{eq:source3}) exhibits several important properties.  The first term of the integrand in Eq.~(\ref{eq:source3}) is resonant for $M_2\sim \mu$ as can be appreciated by rewriting the denominator as \begin{equation} \label{eq:res} \mathcal{E}_{\widetilde{W}}-\mathcal{E}^*_{\widetilde{H}}= \sqrt{ \left| \bf{k}\right|^2+\mu^2+\delta_{\widetilde{H}}^2}-\sqrt{\left| \bf{k}\right|^2+(\mu+\Delta)^2+\delta_{\widetilde{W}}^2}-i(\Gamma_{\widetilde{W}}+\Gamma_{\widetilde{H}}) .\end{equation}  Thus for a given choice of $\mu$ the parameter $\Delta$ determines the strength of the resonance, and hence the resulting baryon asymmetry.   At finite temperature, $\mu(T)$ will generally be different from $\mu(T=0)$, since the singlet vev varies with temperature.  This can be thought of as providing a finite temperature correction to $\Delta$; we neglect this effect in calculating the baryon asymmetry across the parameter space, as this difference depends sensitively on the finite-temperature effective potential at each point.  Note also that the Fermi distribution functions in the numerator result in a suppression of the baryon asymmetry for masses much larger than the electroweak phase transition temperature.  As an optimistic estimate, we take $T_c=140$ GeV in calculating the BAU across the parameter space; the $SU(2)$ sphaleron rate (and hence the overall baryon asymmetry) decreases for lower temperatures.  For example, taking $T_c=100$ GeV will decrease the overall baryon asymmetry by a factor of about 0.7 across the parameter space (i.e. the CP-violating phase $\sin\phi$ at each point would increase by a factor of about 1.4).  We encourage the Reader to bear this in mind while interpreting our results.

Other important quantities determining the strength of the CP-violating source are the bubble wall width ($L_w$), velocity ($v_w$), and the variation of Higgs vevs across the wall ($\Delta \beta$).  This can be seen by approximating the bubble wall profile by a step-function, whence $\dot{\beta}\approx \Delta \beta v_w/L_w$.  For the wall width and velocity we choose $L_w=10/T$ and $v_w=.05$.  These values are inspired by the MSSM and will vary depending on the details of the potential and the spectrum for each point in parameter space as discussed in Sec.~\ref{sec:EWPT}.  In our current set-up, since there is only a small degree of mixing between $A_{1}$ and $A_2$, the quantity $\Delta\beta$ to a good approximation scales as in the MSSM, i.e. roughly $\Delta\beta\propto 1/m_{A_2}^2$ (in our calculation of $\Delta\beta$ we use the full two-loop results of Ref.~\cite{Moreno:1998bq}).  Since $m_{A_2}$ will vary across the parameter space, $\Delta\beta$ will have an important effect on the parameter space available for EWB.  For the values of $m_{A_2}$ we consider, $\Delta\beta$ falls in the range $\Delta\beta\sim 10^{-3}-10^{-4}$. 

The other relevant particle number-changing processes (including the triscalar, Yukawa, and CP-conserving relaxation interactions) are also computed in the vev-insertion approximation; expressions for these rates can be found in Refs.~\cite{Lee:2004we, Chung:2008aya, Lepton_Mediated, Supergauge, Including_Yukawa}.  In addition to these MSSM processes, there are new interactions in the NMSSM arising from the singlet and singlino degrees of freedom.  In particular, there is a resonant relaxation term (and possible CP-violating source \cite{Cheung:2012pg}) arising from higgsino-singlino interactions with the Higgs vevs.  The relevant part of the Lagrangian is \begin{equation} \label{eq:twocomponent} \mathcal{L}_{\rm int}^{\widetilde{S}}=\lambda \left[ v_u(x)\widetilde{H}_d^0 \widetilde{S}+v_d(x) \widetilde{H}_u^0 \widetilde{S}\right] \hspace{.2cm} +\hspace{.2cm}h.c. \end{equation} where $\widetilde{H}^0_{u,d}$ and $\widetilde{S}$ correspond to the two-component  higgsino and singlino fields.  We can rewrite Eq.~(\ref{eq:twocomponent}) in terms of four-component spinors as\begin{equation} \label{eq:fourcomponent} \mathcal{L}_{\rm int}^{\widetilde{S}} = \lambda \bar{\Psi}_{\widetilde{H}^0}\left[ v_u(x)P_L-v_d(x) P_R\right]\Psi_{\widetilde{S}} \hspace{.2cm}+ \hspace{.2cm}h.c.. \end{equation}  and follow the methods of Ref.~\cite{Lee:2004we} to compute the source.  Since we assume that there is no CP-violation in the singlino sector, Eq.~(\ref{eq:fourcomponent}) results in a resonant chiral relaxation rate for the higgsino chemical potential $\Gamma_{\widetilde{H^0}\widetilde{S}}\equiv \Gamma_{\widetilde{H^0}\widetilde{S}}^+ +\Gamma_{\widetilde{H^0}\widetilde{S}}^-$ where \begin{equation} \label{eq:relax} \begin{aligned}  \Gamma_{\widetilde{H^0}\widetilde{S}}^{\pm}=\frac{1}{T}\frac{\lambda^2}{2\pi^2}v(x)^2 \int_0^{\infty}\frac{dk k^2}{\omega_{\widetilde{H}}\omega_{\widetilde{S}}}  &\operatorname{Im} \left\{\left[\mathcal{E}_{\widetilde{S}}\mathcal{E}_{\widetilde{H}}^* -k^2-M_{\widetilde{S}}\left|\mu\right|\sin2\beta\right] \frac{h_F(\mathcal{E}_{\widetilde{S}})\mp h_F(\mathcal{E}_{\widetilde{H}}^*)}{\mathcal{E}_{\widetilde{S}}-\mathcal{E}_{\widetilde{H}}^*}\right. \\ & \left. + \left[\mathcal{E}_{\widetilde{S}}\mathcal{E}_{\widetilde{H}} +k^2+M_{\widetilde{S}}\left|\mu\right|\sin2\beta\right] \frac{h_F(\mathcal{E}_{\widetilde{S}})\mp h_F(\mathcal{E}_{\widetilde{H}})}{\mathcal{E}_{\widetilde{S}}+\mathcal{E}_{\widetilde{H}}} \right\}\end{aligned}\end{equation} and where the various quantities are defined analogously to those in Eq.~(\ref{eq:source3}). The singlino mass given by \begin{equation} \label{eq:singlino} M^2_{\widetilde{S}}=4\kappa^2 \mu^2/\lambda^2 + \delta_{\widetilde{S}}^2\end{equation} (here $\delta_{\widetilde{S}}$ is the singlino thermal mass), and the quantity $h_F$ is defined as \begin{equation} h_F(x)=\frac{e^{x/T}}{\left(e^{x/T}+1\right)^2} .\end{equation} Since we consider moderate values of $\lambda$, we take $\Gamma_{\widetilde{S}}\simeq 0.001 T$ for the singlino width.  The denominator of the first term in Eq.~(\ref{eq:relax}) has the same resonant structure as in Eq.~(\ref{eq:res}) and is the most significant contribution to the transport equations from the singlino, tending to reduce the resulting charge density.  Given our choices for $\lambda$ and $\kappa$ in Fig.~\ref{fig:results}, the relaxation rate $\Gamma_{\widetilde{H}^0\widetilde{S}}$ is near resonance in this region since $M_{\widetilde{S}}\sim \mu$.  We account for this higgsino-singlino resonant relaxation in our computation of the baryon asymmetry, but do not consider the other non-resonant singlet/singlino interactions, as they are subdominant.

\subsection{Solving the Transport Equations}

With the sources contributing to the RHS of Eq.~(\ref{eq:cont}) for the various charged current densities in place, we compute the baryon asymmetry point-by-point across the 130 GeV line parameter space described in Sec.~\ref{subsec:suitable} for $\lambda=0.6$, $\kappa=0.32$, and $\tan\beta=1.8$ as an example.  We do so by solving the system of transport equations to determine the LH charge density $n_L$, assuming a strongly first-order EWPT and that the $SU(2)$ sphaleron rate $\Gamma_{ws}$ is slow compared to the other particle number-changing rates.  Then, given $n_L(z)$, the baryon number density results from the integral of $n_L$ over the unbroken phase, \begin{equation} \label{eq:BAU} n_B= \frac{-3 \Gamma_{ws}}{v_w} \int_{-\infty}^0 dz \hspace{1.5mm} n_L(z)e^{\frac{15\Gamma_{ws}}{4 v_w} z}, \end{equation}  where $z$ is the comoving distance away from the bubble wall (neglecting the curvature of the wall and taking $z<0$ to be the symmetric phase). 

To determine $n_L$, we work under the set of assumptions detailed in Refs.~\cite{Lee:2004we, Kozaczuk:2011vr}, and in particular assuming ``super-equilibrium", i.e. that the chemical potentials of all SM species and their superpartners are equal \cite{Supergauge}.  This allows us to define common charge densities for Higgses and higgsinos, quarks and squarks, etc.  Given the condition of super-equilibrium and that the sfermion masses are heavy, one can show that the relevant charge densities we must keep track of are those corresponding to the Higgs/higgsinos ($H$), the right-handed tops/stops ($T$), and the left-handed third-generation quarks/squarks ($Q$).  The transport equations then read 
 \begin{equation} \begin{aligned}  \label{QBE1} \partial_{\mu} Q^{\mu} = &-\Gamma_{yt}\left(\frac{Q}{k_Q}-\frac{T}{k_T}+\frac{H}{k_H} \right) -\Gamma_{mt} \left(\frac{Q}{k_Q} - \frac{T}{k_T} \right)  - 2\Gamma_{ss} \left( 2\frac{Q}{k_Q}-\frac{T}{k_T}+9\frac{Q+T}{k_B} \right) \end{aligned} \end{equation} \begin{equation} \begin{aligned} \label{QBE2} \partial_{\mu}T^{\mu} = \hspace{.1cm} &\Gamma_{yt}\left(\frac{Q}{k_Q}-\frac{T}{k_T}+\frac{H}{k_H} \right)  +\Gamma_{mt}\left(\frac{Q}{k_Q} - \frac{T}{k_T} \right) +\Gamma_{ss} \left( 2\frac{Q}{k_Q}-\frac{T}{k_T}+9\frac{Q+T}{k_B} \right) \end{aligned} \end{equation}  \begin{equation} \begin{aligned} \label{QBE3} \partial_{\mu} H^{\mu} = &-\Gamma_{yt}\left(\frac{Q}{k_Q}-\frac{T}{k_T}+\frac{H}{k_H} \right) -\Gamma_h \frac{H}{k_H} + S_{\widetilde{H}}^{\slashed{CP}}. \end{aligned} \end{equation} Here, $\Gamma_{mt,h}$ are chiral relaxation rates (including the contribution from the higgsino-singlino-vev interaction), active only in the bubble wall\footnote{For simplicity, in solving the transport equations we assume a step-function profile for the Higgs vevs in the bubble wall.}  and broken EW phase, $\Gamma_{yt}$ are Yukawa interaction rates \cite{Including_Yukawa}, $\Gamma_{ss}$ is the $SU(3)$ sphaleron rate (responsible for generating densities of first- and second-generation quarks), and the $k_i$s are statistical factors relating the charge densities $n_i$ to the corresponding chemical potential $\mu_i$.  We solve Eqs.~(\ref{QBE1})-(\ref{QBE3}) utilizing the diffusion approximation discussed in Ref.~\cite{Lee:2004we}.  The LH charge density entering into Eq.~(\ref{eq:BAU}) is then given to good approximation by the relation \begin{equation} n_L(z)= 5 Q(z)+4T(z). \end{equation}  Contours corresponding to the observed value of the baryon-to-entropy ratio are shown across the 130 GeV line parameter space on the resonance ($\Delta=0$) for different values of the CP-violating phase $\phi$ in Fig.~\ref{fig:results}.  
 
In interpreting our results, the reader should bear in mind that there are several uncertainties present in our calculation of the baryon asymmetry.  As mentioned, the microphysical properties of the EW bubble wall and details of the electroweak phase transition ($L_w$, $v_w$, $\Delta \beta$, $T_c$, etc) can significantly affect the calculation of $n_L$ and $Y_B$ (see e.g.\ Ref.~\cite{Kozaczuk:2011vr} and references therein for a more detailed discussion of these effects).  Also, there are several other frameworks for calculating the baryon asymmetry \cite{CPV_and_EWB, Carena:2002ss, Cline:2000kb, Huber:2001xf, Konstandin:2003dx, Konstandin:2004gy}, with results that can differ by up to an order of magnitude from one another (for a review of these different approaches, see Ref.~\cite{Morrissey:2012db}).  Additionally, there are other possible sources of CP-violation in the NMSSM that could contribute to the BAU in this scenario.  For example, allowing a relative phase between $\lambda$ and $\kappa$ would allow resonant CP-violating singlino sources arising from Eq.~(\ref{eq:fourcomponent}) which in fact would be close to resonant (see Ref.\cite{Cheung:2012pg} for a discussion of singlino-driven EWB in the NMSSM).  

Despite these issues and caveats, Fig.~\ref{fig:results} suggests that resonant CP-violating higgsino-gaugino sources can be very efficient in the region of the NMSSM consistent with a 130 GeV gamma-ray line.  Even if we had over-estimated the baryon asymmetry by an order of magnitude, there could still be regions consistent with both the Fermi line, the observed BAU, constraints from electric dipole moments (which we discuss below), and DM direct detection, provided more optimistic choices for the strange quark content of the proton or the local distribution of dark matter.  For example, taking the values of $\sigma_0$, $\sigma_{\pi N}$ we considered for the EWPT benchmark point pushes out the allowed values of $M_1$ in Fig.~\ref{fig:results} out to about 145 GeV, which would allow a factor of ten over-estimation of the BAU consistent with EDM constraints.

\subsection{EDM Constraints}\label{sec:EDMs}

The NMSSM contains several possible sources of CP-violation beyond those in the MSSM: CP-violation in tree-level parameters $\lambda$, $\kappa$, and $\mu$; CP-violation in soft-breaking terms $A_\lambda$ and $A_\kappa$; and additional effects coming from the mixing between the two CP-odd eigenstates $A_1$ and $A_2$. However, in our setup we assume no CP-violation in the tree-level Higgs sector and very little mixing between $A_1$ and $A_2$ ($A_1$ must be mostly singlet-like, as explained above). Therefore, the electric dipole moment calculations reduce to those in the MSSM.

We use the package {\tt CPSuperH}~\cite{Ellis:2008zy} to calculate the electric dipole moments of the electron, the neutron, and the mercury atom, which have current experimental limits of $|d_e| < 1.05\times10^{-27} e$ cm \cite{Hudson:2011zz} (via the YbF molecule), $|d_n| < 2.9\times 10^{-26} e$ cm \cite{Baker:2006ts}, and $|d_{Hg}| < 3\times10^{-29} e$ cm \cite{Griffith:2009zz}. The neutron and the Mercury atom generally provide extremely strong limits on CP-violating physics, but they are most sensitive to chromo-EDMs and CP-violation involving colored particles. We have no chromo-EDMs in this model, so the electron EDM provides, here, the strongest constraint. All one-loop EDMs are suppressed by the heavy sfermion masses. The dominant two-loop contribution comes from the Barr--Zee diagram containing a chargino loop.

For each point in the parameter space of Fig.~\ref{fig:results}, we calculate the EDMs using the value of $\phi$ that produces the proper baryon abundance. Except for $\phi$, most of the parameters necessary for calculating the EDMs vary little over the plotted region, so the EDMs are most sensitive to $\phi$ and the corresponding iso-level curves follow similar trajectories. The small region in the upper-left with $\sin\phi \gtrsim 0.37$ has $|d_e| > 1.05\times10^{-27} e$ cm, and is thus ruled out by experiment. The smallest EDM in this region, corresponding to $\sin\phi \approx \tfrac{1}{6}$, is $|d_e| =  5.1\times 10^{-28}$. This is well within the anticipated sensitivity of next-generation EDM experiments (for a review, see, {\em e.g.}, Ref.~\cite{Hewett:2012ns}), which have the potential to either rule out or lend credence to this baryogenesis scenario.

\section{Discussion and Conclusions}\label{sec:disc}

The present study reaffirms that the NMSSM framework (and indeed other singlet-extensions of the Higgs sector \cite{SchmidtHoberg:2012ip}) can provide a viable explanation of the 130 GeV Fermi gamma-ray line in terms of resonant neutralino annihilation through a pseudoscalar into photons.  Agreement with observation and with the relevant constraints is realized in the NMSSM for a bino-like LSP (dictating that $M_1\sim 130$ GeV), with relatively large $\lambda$, moderate $\mu$, and with $A_1$ predominantly singlet-like to avoid indirect detection constraints on continuum photons.  While there are many independent constraints on this scenario, currently there remains a substantial amount of parameter space consistent with the gamma-ray line and in agreement with the various dark matter and particle physics constraints.

Here we have shown that the parameter space consistent with the Fermi line in the NMSSM is also promising for electroweak baryogenesis.  In particular, the relatively large values of $\lambda$ typically considered tend to bolster the cubic term in the finite-temperature effective potential in the direction of electroweak symmetry breaking, leading to a strongly first-order electroweak phase transition in parts of the parameter space.  Additionally, the moderate values of $\mu$ ensure that the singlet vev is not too far from the EW scale, again tending towards a strongly first-order transition.  We illustrated this in Sec.~\ref{sec:EWPT} by providing a benchmark point consistent with the 130 GeV line and a strongly first-order EWPT, and in agreement with all other relevant phenomenological constraints.  While we only studied in detail one particular point as a proof of principle, we expect a more systematic study of the NMSSM parameter space to uncover many other regions consistent with the line and a strongly first-order EWPT.

Not only does the parameter space consistent with the line support the possibility of a strongly first-order transition, it can also provide an efficient source for CP-violation that gives rise to the observed baryon-to-entropy ratio of the universe.   Resonant higgsino-gaugino sources can be very efficient here due to the moderate values of $M_{1,2}$ and $\mu$ required to produce the line.  In particular, allowing for a CP-violating phase in $M_2$ does not strongly affect the line or the dark matter phenomenology, but it can produce the observed BAU with $\sin\phi$ small enough to be consistent with electric dipole measurements, as shown in Sec.~\ref{sec:EWB} and Fig.~\ref{fig:results}.  While we focused on the higgsino-wino sources in the present study for the sake of illustration, similar resonant CP-violating sources arising from other interactions can be active in the same regions of parameter space by similar reasoning.  For example, if one allows for $M_1$ to carry a complex phase, resonant bino-higgsino sources can be very efficient as well.  This may be of particular interest in the case of negative $\mu$ whereby $\left|\mu\right|$ can be taken as low as 140 -- 150 GeV (and thus potentially very close to this resonance) while in agreement with direct detection constraints \cite{Chalons:2012xf}.  A careful study of the effect of a CP-violating phase in $M_1$ on the line and dark matter properties would be necessary to assess whether such a scenario is possible, but we expect it is since EDM measurements dictate that the CP-violating phase is necessarily small.  Also, singlino-higgsino sources can in principle be efficient in this region as well, provided a relative phase between $\lambda$ and $\kappa$ \cite{Cheung:2012pg}, again due to the moderate values of the singlino mass (see Eq.~(\ref{eq:singlino})) and $\mu$ in this scenario.  These other sources would be especially important for points such as our EWPT benchmark which features a rather heavy wino but lighter bino and singlino\footnote{Note that non-resonant wino-higgsino sources, such as those considered in Refs.~\cite{CPV_and_EWB, Carena:2002ss, Kozaczuk:2012vx} can also potentially provide the necessary CP-violation for our particular EWPT benchmark point.}. 

An interesting feature of our scenario is that the relevant parameter space will be conclusively tested in the near future by modest improvements in various experimental efforts.  The moderate values of $\mu$ we consider result in rather large spin-independent neutralino-nucleon cross-sections  which continue to be probed by direct detection experiments.  The relatively large values of $\lambda$, as required for a large $\langle \sigma v\rangle_{\gamma \gamma}$, combined with the large $A_{\lambda}$ and moderate values of $\kappa$ necessary for a strongly first-order EWPT, tend towards a significant coupling of $A_1$ to e.g.\ $b \bar{b}$ and so will be tested by modest improvements in indirect detection experiments.  Additionally, the CP-violating phase(s), required to source the left-handed charge density for the $SU(2)$ sphalerons, will be well within reach of various future  EDM experiments (see e.g.\ Ref.~\cite{Kozaczuk:2012xv} for a related discussion). The whole scenario will also continue to be tested by ongoing measurements of Higgs couplings and searches for other particles at the LHC.  

Of course the viability of the 130 GeV line scenario in the NMSSM or any other SM extension hinges on the persistence of the line in the Fermi data and on a dark matter interpretation of these results.  If the line is indeed due to resonant dark matter annihilation, this work shows that the NMSSM framework can potentially explain the origin of both the baryonic and dark matter in our universe. 

\begin{acknowledgments}
\noindent  SP is partly supported by the US Department of Energy under Contract DE-FG02-04ER41268. CLW is supported by an NSF graduate fellowship. 

\end{acknowledgments}

\appendix
\section{Scalar Mass Terms}
\label{sec:scalar_masses}

We present here the scalar mass terms used in the calculation of the finite-temperature effective potential. These are simply the second-derivatives of the full 10-parameter potential, but simplified such that only 3 of the parameters ($h_u$, $h_d$ and $s$) are non-zero. Each subscript denotes a partial derivative with respect to that field. Primed subscripts are derivatives with respect to the imaginary field components, and $\tilde{u}$ and $\tilde{d}$ denote derivatives in the up- and down-type charged directions. The tree-level masses are just the mass eigenvalues of the following matricies.

\begin{align}
M^2_{uu} =& \tfrac{1}{2}\lambda (h_d^2+s^2) + \tfrac{1}{8}(g_1^2+g_2^2) (3h_u^2 - h_d^2) + m_u^2 \\
M^2_{dd} = & \tfrac{1}{2}\lambda (h_u^2+s^2) + \tfrac{1}{8}(g_1^2+g_2^2) (3h_d^2 - h_u^2) + m_d^2 \\
M^2_{ss} = & \tfrac{1}{2}\lambda (h_u^2+h_d^2) + 3\kappa^2 s^2 - \lambda \kappa h_u h_d + m_s^2 + \sqrt{2} \kappa A_\kappa s \\
M^2_{ud} =& \lambda^2 h_u h_d - \tfrac{1}{2} \lambda \kappa s^2 - \tfrac{1}{4}(g_1^2+g_2^2)h_u h_d - \tfrac{1}{\sqrt{2}}\lambda A_\lambda s \\
M^2_{us} =& \lambda^2 h_u s - \lambda \kappa h_d s - \tfrac{1}{\sqrt{2}}\lambda A_\lambda h_d \\
M^2_{ds} =& \lambda^2 h_d s - \lambda \kappa h_u s - \tfrac{1}{\sqrt{2}}\lambda A_\lambda h_u 
\end{align}

\begin{align}
M^2_{u'u'} =& \tfrac{1}{2}\lambda (h_d^2+s^2) + \tfrac{1}{8}(g_1^2+g_2^2) (h_u^2 - h_d^2) + m_u^2 \\
M^2_{d'd'} = & \tfrac{1}{2}\lambda (h_u^2+s^2) + \tfrac{1}{8}(g_1^2+g_2^2) (h_d^2 - h_u^2) + m_d^2 \\
M^2_{s's'} = & \tfrac{1}{2}\lambda (h_u^2+h_d^2) + \kappa^2 s^2 + \lambda \kappa h_u h_d + m_s^2 - \sqrt{2} \kappa A_\kappa s \\
M^2_{u'd'} =& \tfrac{1}{2} \lambda \kappa s^2 + \tfrac{1}{\sqrt{2}}\lambda A_\lambda s \\
M^2_{u's'} =& - \lambda \kappa h_d s + \tfrac{1}{\sqrt{2}}\lambda A_\lambda h_d \\
M^2_{d's'} =& - \lambda \kappa h_u s + \tfrac{1}{\sqrt{2}}\lambda A_\lambda h_u 
\end{align}

\begin{align}
M^2_{\tilde{u}\tilde{u}} =& \tfrac{1}{2} \lambda^2 s^2 + \tfrac{1}{8}(g_1^2+g_2^2) (h_u^2 - h_d^2) + \tfrac{1}{4}g_2^2 h_d^2 + m_u^2 \\
M^2_{\tilde{d}\tilde{d}} =& \tfrac{1}{2} \lambda^2 s^2 + \tfrac{1}{8}(g_1^2+g_2^2) (h_d^2 - h_u^2) + \tfrac{1}{4}g_2^2 h_u^2 + m_d^2 \\
M^2_{\tilde{u}\tilde{d}} =& -\tfrac{1}{2} \lambda^2 h_u h_d + \tfrac{1}{2} \lambda \kappa s^2 + \tfrac{1}{4} g_2^2 h_u h_d + \tfrac{1}{\sqrt{2}}\lambda A_\lambda s
\end{align}

There is a second matrix for the charged Higgs, but the two are identical except for a change of sign in the off-diagonal term which does not affect its eigenvalues.

In the high-temperature approximation, the thermal mass terms come from the quadratic piece of the one-loop finite-temperature contributions to the effective potential. The scalar thermal masses include contributions from all particles with field dependent masses, and they get added to each of the diagonal terms in the mass matrix. They are:

\begin{align}
\Pi_u &= T^2\left[ \tfrac{1}{8}\left(g_1^2 + 3g_2^2 \right) + \tfrac{1}{4}\lambda^2 + \tfrac{1}{4}y_t^2 \right] \\
\Pi_d &= T^2\left[ \tfrac{1}{8}\left(g_1^2 + 3g_2^2 \right) + \tfrac{1}{4}\lambda^2 + \tfrac{1}{4}y_b^2 \right] \\
\Pi_s &=  \tfrac{1}{2} T^2 \left( \lambda^2 + \kappa^2 \right).
\end{align}

The longitudinal polarizations of the gauge bosons also receive thermal mass corrections. At finite temperature, the gauge boson mass mixing is
\begin{align}
M^2_{gauge-long} = \frac{h_u^2 +h_d^2}{4}\begin{pmatrix}
g_2^2  & & & \\
& g_2^2 & & \\
& & g_2^2 & g_1 g_2 \\
& & g_1 g_2 & g_1^2
\end{pmatrix} +
T^2\begin{pmatrix}
\tfrac{5}{2}g_2^2  & & & \\
& \tfrac{5}{2}g_2^2  & & \\
& & \tfrac{5}{2}g_2^2  &  \\
& & & \tfrac{13}{6} g_1^2
\end{pmatrix}.
\end{align}
Again, we have ignored the contributions from the sfermions, because they are much too heavy to factor into the high-temperature corrections. For more information on thermal masses in the supersymmetric theories, see ref.~\cite{Comelli:1996vm}.

\end{document}